\documentclass{aa}  
\usepackage{natbib}

\usepackage{graphicx}
\usepackage{txfonts}
\newcommand{\logg}{\ensuremath{\log g}}

\newcommand{\ferre}{{\tt FER\reflectbox{R}E}}
\begin{document} 

   \title{WHT follow-up observations of extremely metal-poor stars identified from SDSS and LAMOST}

   %\subtitle{}

      \author{D.~S. Aguado\inst{1,2}, J.~I. Gonz\'alez Hern\'andez\inst{1,2}, C. Allende Prieto\inst{1,2}, R. Rebolo\inst{1,2,3}}
      %, Rafael Rebolo\inst{1,2,3}}

   \institute{Instituto de Astrof\'{\i}sica de Canarias,
              V\'{\i}a L\'actea, 38205 La Laguna, Tenerife, Spain\\              
         \and
             Universidad de La Laguna, Departamento de Astrof\'{\i}sica, 
             38206 La Laguna, Tenerife, Spain \\  
         \and
             Consejo Superior de Investigaciones Cient\'{\i}ficas, 28006 Madrid, 
             Spain\\
            }   
  
\authorrunning{D.S. Aguado et al.}
\titlerunning{EMP stars identified from SDSS and LAMOST}

% \abstract{}{}{}{}{} 
% 5 {} token are mandatory
 
  \abstract
   {}
   {We have identified several tens of extremely metal-poor star 
   candidates from SDSS and LAMOST, which we follow up with 
   the 4.2m William Herschel Telescope (WHT) telescope to confirm their metallicity.}
   {We followed a robust two-step methodology. We first analyzed the 
   SDSS and LAMOST spectra.
   A first set of stellar parameters was derived from these spectra 
   with the FERRE code, taking advantage of the continuum shape 
   to determine the atmospheric parameters, in particular, the effective 
   temperature.   
   Second, we selected interesting targets for follow-up observations, some 
   of them with very low-quality SDSS or LAMOST data. We then 
   obtained and analyzed higher-quality medium-resolution
   spectra obtained with the Intermediate dispersión Spectrograph and Imaging System (ISIS) on the WHT to arrive at  
   a second more reliable set of atmospheric parameters. 
   This allowed us to derive the metallicity with accuracy, and we confirm the extremely metal-poor nature in most cases. 
   In this second step we also employed FERRE, but we took a 
   running mean to normalize both the observed and the synthetic 
   spectra, and therefore 
   the final parameters do not rely on having an accurate 
   flux calibration or continuum placement. We have analyzed with the same tools 
   and following the same procedure six well-known metal-poor stars, 
   five of them at [Fe/H]$<-4$ to verify our results. This showed that our methodology is 
   able to derive accurate metallicity determinations down to [Fe/H]$<-5.0$.
   }
   {The results for these six reference stars give us confidence 
   on the metallicity scale for the rest of the sample. In addition,
   we present 12 new extremely metal-poor candidates: 
   2 stars at [Fe/H] $\simeq -4$, 6 more 
   in the range $-4<\left[{\rm Fe/H}\right]<-3.5$, and 4
   more at $-3.5<\left[{\rm Fe/H}\right]<-3.0$.}
   {We conclude that we 
   can reliably determine metallicities
   for extremely metal-poor stars with a precision of 0.2\,dex from
   medium-resolution spectroscopy with our improved methodology. This provides
   a highly effective way of verifying candidates from lower quality data.
   Our model spectra and the details of the fitting
   algorithm are made public to facilitate the standardization of the analysis 
   of spectra from the same or similar instruments.}
   

   \maketitle
%
%-------------------------------------------------------------------

\begin{table*}
\begin{center}
 \renewcommand{\tabcolsep}{5pt}
\centering

\caption{Coordinates and atmospheric parameters for the program 
stars based in the analysis of the low-resolution spectra with FERRE.
\label{basic}}
\begin{tabular}{lccccccccccc}
\hline
Star & $g$ &      RA &      DEC &$T_{\rm eff}$ & $\log g$  & $\left[{\rm Fe/H}\right]$ 
& $\left[{\rm C/Fe}\right]$ &$\rm <S/N>^{a}$&survey\\
     & mag & h  '   ''&$\mathring{}$  '  ''   & K & $\rm cm\, s^{-2}$ &               &                       \\
\hline\hline
SDSS J015131+163944    & 18.9 & 01:51:31.2 &  +16:39:44.98&5917 &3.9&-3.6 &1.3 & 19& BOSS \\
SDSS J030444+391021    & 17.8& 03:04:44.97 & +39:10:21.17&  5918&4.9&-3.7 &  0.3 &29&SEGUE  \\
SDSS J105519+232234    & 17.8&  10:55:19.28 & +23:22:34.03& 6448&4.9& -4.2 &-0.3 & 32&BOSS  \\
SDSS J132917+542027    & 14.9 &13:29:17.34 & +54:20:27.52 &5876 &0.5&  -4.0& $-$& 47&LAMOST  \\
SDSS J134157+513534    & 15.4&13:41:57.97 & +51:35:34.08 &6011 &1.5&-4.9 & $-$ &  35&LAMOST \\
SDSS J134338+484426    & 12.6 &13:43:38.66 & +48:44:26.48 &5390 & 1.0&-3.7 &0.7 &56&SEGUE   \\
SDSS J152202+305526    & 16.6 &15:22:02.09 & +30:55:26.29 &5500 &0.6& -4.1& $-$  & 37&SEGUE  \\ 
SDSS J173329+332941    & 18.9 & 17:33:29.32 & +33:29:41.94& 6448&4.5&-4.2 & 0.7 &21& SEGUE  \\
SDSS J200513$-$104503  & 17.0 & 20:05:13.48 & -10:45:03.21 &6488 & 4.4&-3.7 &-0.7 &45&SEGUE  \\
SDSS J202109+601605    & 17.9 & 20:21:09.01&  +60:16:05.33&  5834& 3.3& -3.4& 0.4& 20&SEGUE  \\ 
SDSS J204524+150825    & 16.7& 20:45:24.03 & +15:08:25.46&5069 &2.4&-3.6 & 0.0 & 39&SEGUE  \\
SDSS J231027+231003    & 17.3 &23:10:27.16 & +23:10:03.43 &6188 &4.1&-4.0 & $-$ &33&SEGUE  \\ 
SDSS J222505+164322    & 18.0 & 22:25:05.97 & +16:43:22.52&5378 &1.0&-3.7 &0.9&24&BOSS  \\ 
\hline
\hline
\end{tabular}
\end{center}

 \tablefoot{ $^a$ Signal-no-noise ratio have been calculated as average of the SDSS or LAMOST entire spectrum}

\end{table*}
\section{Introduction}\label{intro}

The oldest stars in the Galactic halo contain information about the early
Universe after the primordial nucleosynthesis. These stars
are key for understanding how galaxies form, what the masses of the first 
generation of stars were, and what the early chemical evolution of the Milky Way was like.
The oldest stars are extremely poor in heavy elements 
and havemost likely been preceded locally by only one massive Population III star
\citep{fre15}. 
Chemical abundances in extremely metal-poor (EMP) stars (with
$\left[{\rm Fe/H}\right]<-3$), in particular dwarf EMP stars, are needed
to study the difference between the primordial lithium abundance obtained from
standard Big Bang nucleosynthesis, constrained by
the baryonic density inferred from the Wilkinson Microwave Anisotropy Probe (WMAP) observations of the Cosmic microware background (CMB) \citep{spe03},
and 
in situ atmospheric abundance measurements from old  
metal-poor turn-off stars \citep{spi82,reb88,gon08,boni09,sbo12}.
Unfortunately,  very metal-poor stars are extremely rare, with fewer than ten 
stars known in the $\left[{\rm Fe/H}\right]<-4.5$ regime. This seriously limits the number of lithium measurements and detections that are
required to shed light on this problem.

Detailed chemical abundance determinations of EMP stars require high-resolution spectroscopy \citep[e.g.][]{roe14,frerev2}, but the identification
of these stars is a difficult task.
In the 1980's and 1990's several different techniques were designed
and applied to search for metal-poor stars of the Galactic halo
using various telescopes that were equipped with low- and medium-resolution
spectrographs~\citep{bee85, bee92, rya91, car96}. 
More recently, the stars with the lowest iron abundances have been discovered 
from candidates identified in the Hamburg-ESO survey \citep{chris02} or the Sloan 
Digital Sky Survey \citep[SDSS][]{yor00, boni12}.
In the past decade, additional stars have been 
identified by photometric surveys such as the one using the SkyMapper
Telescope, see \cite{kell12}, or the Pristine Survey \citep{sta17}, 
and spectroscopic surveys such as the Radial Velocity Experiment (RAVE,
\citealt{ful10}), the Sloan Extension for Galactic Understanding and
Exploration (SEGUE,\citealt{yan09}), or the large Sky Area 
Multi-Object Fiber Spectroscopic Telescope (LAMOST, \citealt{deng12}).
 Extremely metal-poor stars tend to be located at large distances. 
 For these stars, spectroscopic surveys usually do not provide spectra 
 with sufficient quality, and it is therefore critical 
 to examine and verify candidates with additional observations with  
higher quality. 

 In this work, we follow a two-step methodology to identify new
 extremely metal-poor stars:
 We perform an improved analysis of SDSS and LAMOST data
 to select EMP candidates, and we carry out  follow-up observations of a  subsample, which allows us to check the stellar parameters, metallicities, and 
carbon abundances of these candidates. 
 The paper is organized as follows.
In Section~\ref{selection} we explain the candidate selection. 
Section \ref{obs} is devoted to the follow-up observations and data reduction.
Section~\ref{anal} describes in detail how the 
analysis was carried out, including tests using well-known metal-poor stars 
(Section~\ref{known}). 
We then discuss our carbon abundance determination from low- and 
medium-resolution spectra (Section~\ref{carbon}). 
Finally, Section~\ref{discuss} summarizes our results and conclusions.

%--------------------------------------------------------------------

\section{Low spectral resolution analysis and target selection}\label{selection}

We have analyzed the low-resolution spectra of more than 2.5 million 
targets from three different surveys:  the Baryonic Oscillations 
Spectroscopic Survey (BOSS, \citealt{eis11,daw13}), SEGUE \citep{yan09}, 
and LAMOST \citep{deng12}. 
The analysis of the stars from these surveys was performed  
with \ferre\ \footnote{{\tt FER\reflectbox{R}E} is available 
from http://github.com/callendeprieto/ferre} \citep[]{alle06}, which allowed us 
to derive the main stellar parameters, effective temperature $T_{\rm eff}$, 
and surface gravity $\log g$, together with the metallicity [Fe/H] 
\footnote{We use the bracket notationto report
chemical abundances: [a/b]$ = \log \left( \frac{\rm N(a)}{\rm N(b)}\right) - \log
\left( \frac{\rm N(a)}{\rm N(b)}\right)_{\odot}$,
where $\rm N$(x) represents number the density of nuclei of the element x.}  
and carbon abundance [C/Fe] \citep[see][for further details]{agu16}. 
As a result of the extremely low metal abundances of our targets,  
iron transitions are not detected, either individually
or even as blended features, in medium-resolution spectra. 
We typically measured only the \ion{Ca}{ii} resonance lines and used them as 
a proxy for the overall metal abundance of the stars, assuming the typical 
iron-to-calcium ratio found in metal-poor stars, [Fe/H]=[Ca/H]$-0.4$. 

We analyzed about 1,700,000 ($\sim$1,500,000 stars) spectra 
from LAMOST, 740,000 ($\sim$660,000 stars) from 
SEGUE and 340,000 ($\sim$300,000 stars) from BOSS.
A database with all candidates at [Fe/H]$<-3.5$ was 
built that contains $\sim500$ objects with a first set of stellar properties
($T_{\rm eff}$, $\log g$, [Fe/H], and [C/Fe]). 
However, this first selection still contains many outliers and
unreliable fits. 
We reevaluated the goodness of fit by
measuring the $\chi^{2}$ in a limited spectral region around the Balmer lines.  
This method prevents false positives even at low signal-to-noise ratios
(S/N<20).
Finally, a visual inspection of the spectra and their best fittings 
was carried out to select, to skim the best several tens of candidates.
The stars that passed this selection were then scheduled for 
observations at medium-resolution and much higher S/N ratio.
Table~\ref{basic} shows the derived parameters of the final sample that was analyzed 
in this work.
Visual inspection helps us to detect promising candidates. 
The two LAMOST objects discussed below are good example for cases where 
visual inspection was critical.
In addition to these stars, we chose some of the most 
metal-poor stars known ([Fe/H]$<-4.0$), with published determinations 
from high-resolution spectroscopic data, in order to test our methodology.

%--------------------------------------------------------------------

\section{Observations and data reduction}\label{obs}

The second step in our methodology makes use of follow-up medium-resolution spectroscopy obtained with the Intermediate dispersion Spectrograph and Imaging System (ISIS) \citep{isiswht} spectrograph 
on the 4.2\,m William Herschel Telescope (WHT) at the Observatorio del 
Roque de los Muchachos (La Palma, Spain).
We used the R600B and R600R gratings, the GG495 filter in the red arm, 
and the default dichroic (5300\,\AA). 
The mean FWHM resolution with a one-arcsecond slit was R$\sim 2400$ in 
the blue arm and R$\sim 5200$ in the red arm.
More details are provided in \citet{agu16}.
The observations were carried out over the course of five observing runs: 
run I: Dec 31 - Jan 2, 2015 (three nights); run II: February 5-8, 2015 (four nights); run III: 
August 14-18,2015 (five nights); run IV: May 1-2, 2016 (two nights), and run V: July 
29 - 31, 2016 (three nights). 
A standard data reduction (bias substraction, flat-fielding and wavelength 
calibration, using CuNe $+$ CuAr lamps) was performed with the 
\emph{onespec} package in IRAF\footnote{IRAF is distributed by the 
National Optical Astronomy Observatory, which is operated by the 
Association of Universities for Research in Astronomy 
(AURA) under cooperative agreement with the National Science 
Foundation} \citep{tod93}.

High-resolution spectra of well-known metal-poor stars were 
extracted from the ESO archive\footnote{Based on data from the 
ESO Science Archive Facility %under request number 
%SAF 158898, SAF 164850, SAF 183152, 
%PHASE3 186337, PHASE3 226010, PHASE3 226022, 
%PHASE3 229141, PHASE3 229144, PHASE3  229145, 
%PHASE3 229146, PHASE3 248261, PHASE3 249503,
%PHASE3 249504, SAF 250238
} and obtained with the Ultraviolet and 
Visual Echelle Spectrograph (UVES, \citealt{dek00}) at the VLT. 
The spectral range of these data is 3300-4500\,\AA\ and the resolving 
power is about 47,000. 
The spectra were reduced using the automatic UVES pipeline
\citep{ball00}, offered to the community through the ESO Science 
Archive Facility.

\begin{figure}
\begin{center}
{\includegraphics[width=90 mm]{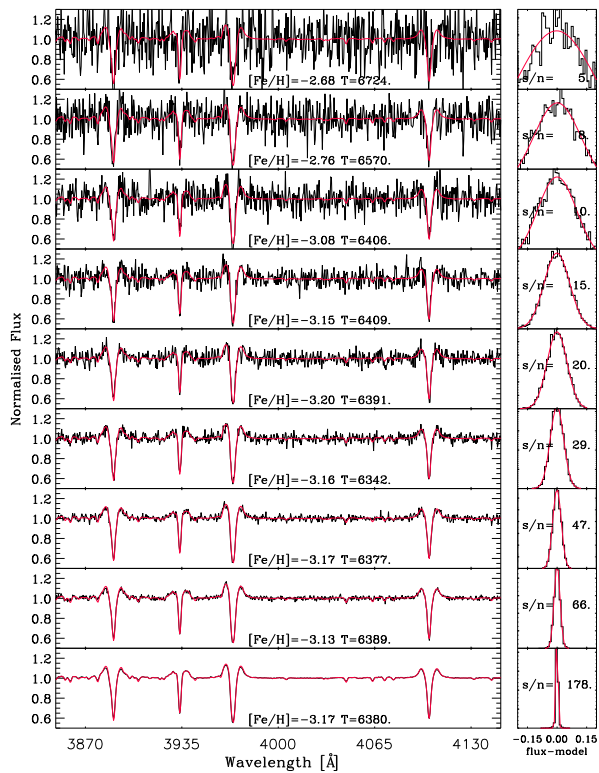}}
\end{center}
\caption{Left panel: medium-resolution ISIS spectrum of G64-12 with 
different levels of added Poisson noise (black lines) and the best fit 
computed with FERRE (red line). 
Right panel: Gaussian distribution of the S/N 
computed on the residuals from subtracting the best-fitting model from
the observed spectrum.}
\label{sta}
\end{figure}
 \begin{figure}
\begin{center}
{\includegraphics[width=90 mm, angle=180]{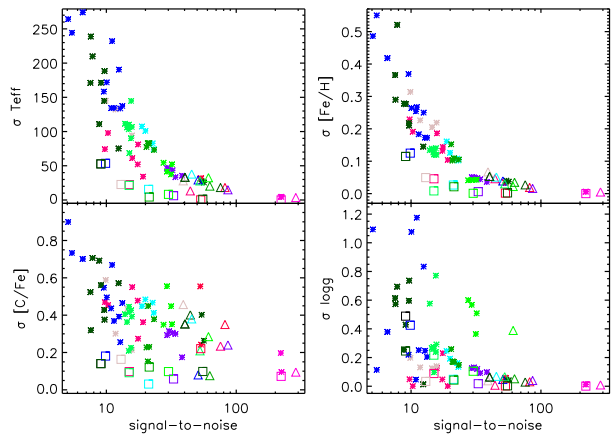}}
\end{center}
\caption{Random errors derived with FERRE as a function of the 
S/N for individual and combined exposures.
Each color represents a different star. 
The triangles correspond to the combined spectra, and  
asterisks indicate the individual exposures, with random errors 
estimated with FERRE and given in Table \ref{AnalysisResults}.
The squares show the dispersion in the parameters derived 
from individual exposures, shown at the mean value of their S/N.}  
\label{err}
\end{figure}

%--------------------------------------------------------------------

\section{Analysis and discussion}\label{anal}

In order to derive the stellar parameters and chemical abundances a 
grid of synthetic 
spectra was computed with the ASS$\epsilon$T code 
\citep{koe08}, which uses the Barklem codes \citep{ber00c, ber00d}
to describe the broadening of the Balmer lines. 
This grid resembles the grid used by \citet{alle14}, with some 
differences described below, and it is provided ready to be
used with \ferre\, as an electronic table. 
The model atmospheres were computed with the same Kurucz 
codes and  strategies as described by \citet{mez12}. 
The abundance of $\alpha-$elements was fixed to $[\alpha/\rm Fe]=+0.4$,
and the limits of the grid were $-6\leq[\rm Fe/H]\leq-2$; $-1 \leq[\rm C/Fe]\leq 5$, 
$4750\,\rm K\leq T_{eff}\leq7000\,\rm K$ and $1.0\leq \logg \leq5.0$,
assuming a microturbulence of 2\,km s$^{-1}$.
We searched for the best fit using \ferre\, by simultaneously deriving
the main three atmospheric parameters and the carbon abundance.

Interstellar reddening and the instrument response distort the intrinsic
shape of the stellar spectra.
To minimize systematic differences between the synthetic spectra and 
the observations,  we worked with continuum-normalized fluxes.
The observed and synthetic spectra were both normalized using a
running-mean filter with a width of 30 pixels (about 10\,\AA).
Although the precision in the derived parameters was 
found to be very sensitive to the normalization scheme, our tests 
showed that 
%in the most general case, 
when no reliable information 
on the shape spectral energy distribution is available, 
the running-mean filter offers very good performance. 
With this modification we were able to retrieve  
$T_{\rm eff}$ and $\logg$ information from ISIS spectra.

In order to determine how robust the \ferre\ results are, we 
carried out a series of consistency tests by adding random noise with 
a normal distribution to an observed spectrum of the metal-poor star 
G64-12. The results are illustrated in Fig. \ref{sta} and are described
below.
We find that \ferre\ is able to recover the original set of parameters 
from high S/N values down to S/N~$=15$ when the derived solution 
begins to drift away from the expected solution. 
This is consistent with the results obtained by \citet{alle14} 
for SDSS spectroscopy.

We used observations of well-known metal-poor stars to test our methodology. 
The comparison between the literature metallicities 
from high-resolution analyses and those we obtained  
using medium-resolution spectroscopy indicates that our 
approach works quite well.
Nonetheless, our methodology does not allow us to separate the 
\ion{Ca}{ii} K spectral lines from the star from the interstellar medium (ISM) contributions that are 
not resolved in the ISIS spectra. 
Below, we describe tests in which we reanalyzde the original 
UVES spectra (see Section~\ref{known}) of several objects after smoothing
to understand the effect of having a lower spectral resolution.
 
\ferre\ is able to derive internal uncertainties in several ways. After 
extensive testing in our ISIS spectra, we find that the best option 
is to inject noise using a normal distribution to the observed spectra 
several times, and search for the best-fitting parameters. 
Such Monte Carlo simulations were performed 50 times for each 
observed spectrum. 
The dispersion in the derived parameters was adopted as a measurement
of the internal uncertainty on these parameters in our analysis.
In addition to the final combined spectra for each target, we
analyzed each individual exposure in the same way.
Figure \ref{err} shows the relationship between our derived internal 
random uncertainties and the S/N. 
%The asterisks represent the individual spectra of each star, the 
%triangles are the combined spectra and 
%the squares are the average
%of the individual spectra for each star. 
The uncertainties on the final combined spectra follow 
the same trend as was found in the analysis of individual exposures.
The only giant star in the sample, J2045+1508, shows a higher
uncertainty in its surface gravity determination, as expected
(see Table \ref{AnalysisResults}).
From these tests, we conclude that our random errors are 
consistently and reliably determined, with the exception of the 
[C/Fe] abundance ratios, for which we tend to underestimate
the random uncertainties, given that the dispersion from individual
estimates of this parameter is larger than the random error 
from individual exposures. 
This is particularly relevant at [C/Fe]$<1.0$ dex.

In addition to random errors, other sources of systematic uncertainty 
can affect our results.
Stellar effective temperatures are not easy to determine accurately 
\citep[see, e.g.,][]{sbo10}. 
We adopted a 100\,K uncertainty following the comparison between
photometric and spectroscopic methods by \citet{yong04}.
In addition, the stellar isochrones depicted in Fig. \ref{iso} suggest that  
systematic $0.3-0.5$ dex error could be present in our spectroscopic 
gravities (see Sects. \ref{new} and \ref{discuss}). 
Based on these comparisons, we decided to adopt a systematic uncertainty 
in surface gravity of 0.5 dex.
The error bars shown for the data in Fig. \ref{iso} are the quadratic 
combination of our estimated random and systematic uncertainties.
We adopted a 0.1\,dex metallicity error, according to the comparisons 
between our results and those in the high-resolution analyses in 
the literature, but we caution that much larger systematic
uncertainties could be present as a result of departures from local thermal equilibrium (LTE) and 
hydrostatic equilibrium (see \citet{schu05,nor17} and references therein), 
which could add up to 0.8-0.9\,dex.
Based on other works (see Section \ref{carbon}) we estimate that our 
derived carbon abundances could have a systematic uncertainty of 
about 0.3\,dex.
In Table \ref{car_tab} we show the carbon abundances for well-known 
metal-poor stars whose uncertainties were computed by adding in 
quadrature the random error and the adopted systematic error.

%--------------------------------------------------------------------

\subsection{Well-known metal-poor stars}\label{known}

\subsubsection{G64-12}\label{g64-12}
\citet{aoki06I} observed G64-12 with the High Dispersion 
Spectrograph (HDS) at the Subaru telescope and derived its 
main atmospheric parameters as 
$T_{\rm eff}=6390\pm 100\,K$, $\logg=4.4\pm 0.3$, and
$ \left[{\rm Fe/H}\right]=-3.2\pm 0.1$.
We obtained with ISIS at the WHT a high-quality spectrum (S/N$\sim300$) of 
this object and derived a set of parameters that excellently agree 
with those from \citet{aoki06I}: 
$T_{\rm eff}=6377\pm 104\,K$, $\logg=4.8\pm 0.7$, and 
$\left[{\rm Fe/H}\right]=-3.2\pm 0.2$.

\subsubsection{SDSS J1313$-$0019}\label{allende}

There is an open debate about the metallicity of this object, discovered by 
\cite{alle15} using a low-resolution spectrum from the BOSS project 
\citep{eis11,daw13}. \cite{alle15} derived the following set of parameters: 
$T_{\rm eff}=5300\pm 50\,K$, $\logg=3.0\pm 0.2$, 
$\left[{\rm Fe/H}\right]=-4.3\pm 0.1$, and $\left[{\rm C/Fe}\right]=2.5\pm 0.1$. 
This temperature reproduces both the local continuum slope and the shape of the 
Balmer lines in the BOSS spectrum of the star. 
Shortly after the discovery, \cite{fre15} obtained a high-resolution ($R \sim 35,000$) 
spectrum with the MIKE spectrograph at the Magellan-Clay telescope.
Their adopted effective temperature and surface gravity  
are $T_{\rm eff}=5200\pm 150\,K$ and $\logg=2.6\pm 0.7$. 
They measured 37 \ion{Fe}{I} lines (\ion{Fe}{II} features were not detected) 
and estimated a metallicity of $\left[{\rm Fe/H}\right]=-5.00\pm 0.28$ and
$\left[{\rm C/Fe}\right]=2.96\pm 0.28$.

\begin{figure}
\begin{center}
{\includegraphics[width=100 mm, angle =180]{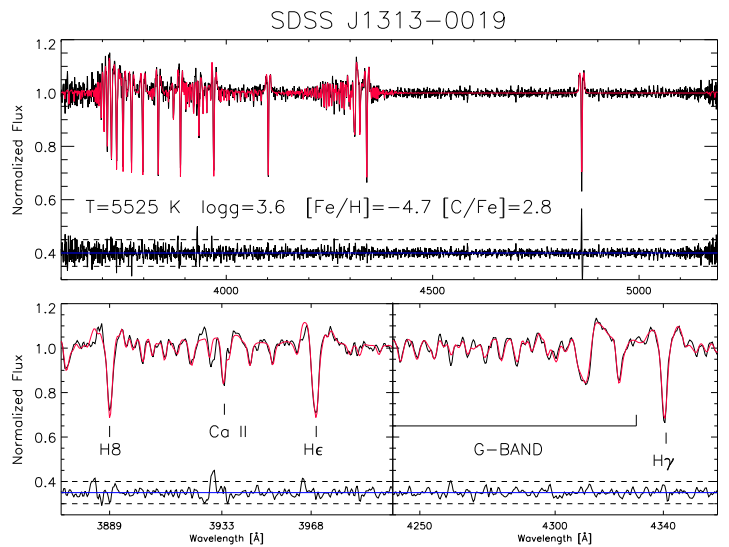}}
\end{center}
\caption{Top panel: blue ISIS J1313$-$0019 spectrum (black line) 
and the best fit computed with FERRE (red line) with the residuals (black line 
over blue reference). 
The derived stellar parameters and carbon abundance are shown. 
Bottom panel: detail of the \ion{Ca} {II} H-K and the g-band spectral region.
}
\label{j1313}
\end{figure}

A high-S/N medium-resolution spectrum of J1313$-$0019 was 
obtained as part of our WHT program.
Figure~\ref{j1313} shows the entire spectrum and the best-fitting spectrum from the 
\ferre\ analysis.
We derive $T_{\rm eff}=5525\pm 106\,K$, $\logg=3.6\pm 0.5$, 
$\left[{\rm Fe/H}\right]=-4.7\pm 0.2$, and $\left[{\rm C/Fe}\right]=2.8\pm 0.30$. 
Our derived metallicity is in between those by \citet{alle15} 
and \citet{fre15}, but closer to the determination by \citet{fre15}.
When we use carbon-enhanced model atmospheres, consistent with the high 
[C/Fe] ratio adopted in the spectral synthesis, it does not introduce 
significant changes in the derived values.
The difference between our analysis and that in \citet{alle15} is 
partly explained by the fact that we assume $[\rm \alpha/H]=0.4$, while 
the authors derived $[\rm \alpha/H]=0.2\pm 0.1$.

We checked that when we impose in the \ferre\ analysis 
the same effective temperature and surface gravity as adopted 
by \citet{fre15} ($T_{\rm eff}=5200$ and  $\logg=2.6$), we recover 
a metallicity and carbon abundance of $\left[{\rm Fe/H}\right]=-4.9\pm 0.2$ 
and $\left[{\rm C/Fe}\right]=2.7\pm 0.3$, respectively. 
The adoption of an effective temperature 300 K cooler by \citet{fre15} 
is responsible for an offset of about 0.2\,dex in metallicity.  
Since our value of $T_{\rm eff}$ (see Table~\ref{AnalysisResults})
is mostly based on fitting all the Balmer lines available in the spectrum 
we consider it reliable.

\subsubsection{HE 0233$-$0343}\label{hansen}

This star was studied by \citet{han14} using high-resolution spectra
acquired with UVES on the VLT, and it is one of the rare stars
at such a low metallicity where the lithium abundance was 
measured, $\rm A(Li)=1.77$.
These objects are key for understanding the cosmological Li 
problem~\citep[e.g.][]{asp06,boni07,gon08}.
Our effective temperature determination, $T_{\rm eff}=6150\pm 103\,K$,
is consistent with the temperature proposed by \citet{han14} of $T_{\rm eff}=6100\pm 100\,K$.
They assumed an age of 10 Gyr to infer a surface gravity of $\logg=3.4\pm 0.3$  
from isochrones, while we derive $\logg=4.9\pm 0.7$ from the ISIS spectrum.

Calcium ISM absorption is clearly visible in the UVES spectrum 
of this star, but it is not resolved at the ISIS resolution (see Fig. \ref{comp}).
An unresolved ISM contribution biases our 
metallicity determination from
 \ferre\ to $\left[{\rm Fe/H}\right]=-4.0\pm 0.1$, 
while \citet{han14} derived $\left[{\rm Fe/H}\right]=-4.7\pm 0.2$.
To check the consistency of the high- and medium-resolution results,  
we smoothed the UVES spectrum to the ISIS resolution and 
reanalyzed it. We arrived at the same result as for the ISIS data:
$T_{\rm eff}=6207\,K$; $\logg=4.9$, and $\left[{\rm Fe/H}\right]=-4.0$, 
as illustrated in Fig.~\ref{comp}.

\subsubsection{SDSS J1029+1729}\label{caffau}

The star SDSS J1029+1729 has no carbon (or nitrogen) detected, which 
makes it the most metal-poor star ever discovered~\citep{caff11,caff12I}. 
This rare object challenges theoretical calculations that predict that no 
low-mass stars can form at very low metallicities. 
Using UVES with a resolving power $R\sim38000$,
\citep{caff12I} derived the following set of parameters: 
$T_{\rm eff}=5811\pm 150\,K$, $\logg=4.0\pm 0.5$, and 
$\left[{\rm Fe/H}\right]=-4.9\pm 0.2$. 
Our analysis of the medium-resolution ISIS spectra arrives at 
$T_{\rm eff}=5845\pm 105\,K$, $\logg=5.0\pm 0.8$ and 
$\left[{\rm Fe/H}\right]=-4.4\pm 0.2$. Following a similar argument 
as given in Section~\ref{hansen}, the difference in metallicity is associated
with the calcium ISM contribution (See Fig.~\ref{comp}). 
This is demonstrated by degrading the UVES spectrum to the resolution 
of the ISIS data. 
This analysis gives a result that is fully consistent with the result from 
the ISIS spectrum: 
$T_{\rm eff}=5867\,K$, $\logg=5.0$, and $\left[{\rm Fe/H}\right]=-4.5$.

\subsubsection{HE 1327$-$2326}\label{frebel}

The study of this star is complicated because of the complex structure of 
the calcium ISM features and the extraordinary amount of carbon 
($\rm [C/Fe]=4.26$, \citealt{fre05,aoki06I}) in the stellar 
atmosphere (see Fig.~\ref{comp}).
The parameters from \citet{aoki06I} from a high-resolution UVES 
spectrum are $T_{\rm eff}=6180\pm 100\,K$, $\logg=3.7\pm 0.3$, and 
$\left[{\rm Fe/H}\right]=-5.6\pm 0.1$ while \ferre\ returns 
$T_{\rm eff}=6150\pm 102\,K$, $\logg=4.3\pm 0.7$, and  
$\left[{\rm Fe/H}\right]=-4.8\pm 0.2$ from the medium-resolution ISIS spectrum. 
We display in  Fig.~\ref{comp} the original UVES observation, 
the same data smoothed to the resolution of ISIS, and the ISIS spectrum.
 
The UVES spectrum resolves multiple ISM features, while ISIS only uncovers 
some of the blue components so that the \ferre\ analysis of the ISIS spectrum 
returns a higher metallicity. 
The smoothed spectrum solution is again consistent with our 
own medium-resolution ISIS values:
$T_{\rm eff}=6119\,K$; $\logg=4.3,$ and $\left[{\rm Fe/H}\right]=-4.9$. The 
use of medium-resolution data inevitably leads to a biased metallicity because of the 
unresolved ISM calcium absorption.

\begin{table*}
\begin{center}
\renewcommand{\tabcolsep}{5pt}
\centering
\caption{Stellar parameters and main results obtained from ISIS spectra.
\label{AnalysisResults}}
\begin{tabular}{lcccccccccccc}
\hline
Star              & $\rm N_{exp}$& $\langle\rm S/N\rangle^a$ & ${\rm T_{\rm eff}}$ &$\Delta\rm T_{\rm eff}$  &\logg &$\Delta \logg$  &  $[\rm Fe/H]$ & $\Delta[\rm Fe/H] $ &$[\rm C/Fe]$ &$\Delta[\rm C/Fe] $\\
                    &1800\,s &  & [K] & [K]& [cm\, s$^{-2}$] & [cm\, s$^{-2}$]& & \\

\hline\hline
SDSS J015131+163944  & 10&40  &6032&33   &4.8&0.07 &-3.8&0.05 & 1.3  &0.35\\  
SDSS J030444+391021  & 6 &44  &5859&13   &5.0&0.01 &-4.0&0.05 & <0.7 & --\\  
SDSS J105519+232234  & 4 &39  &6232&28   &4.9&0.03 &-4.0&0.07 & <0.7 &--\\  
SDSS J132917+542027  & 2 &82  &6322&18   &4.7&0.03 &-2.7&0.02 & <0.7  &--\\  
SDSS J134157+513534  & 2 &76  &6358&18   &4.8&0.02 &-3.5&0.03 & <0.7 &-- \\ 
SDSS J134338+484426  & 2 &287 &5888&4   &4.1&0.01 &-3.3&0.01 & 0.8  &0.09\\  
SDSS J152202+305526  & 6 &86  &5865&15   &4.0&0.04 &-3.7&0.02 & 0.8  &0.24\\  
SDSS J173329+332941  & 12&51  &6325&29   &4.7&0.05 &-3.4&0.04 & 1.8  &0.08 \\ 
SDSS J200513$-$104503& 4 &45  &6361&38   &4.6&0.06 &-3.5&0.05 &<0.7  &--\\  
SDSS J202109+601605  & 9 &52  &5973&28   &4.7&0.07 &-3.4&0.03 & 0.7  &0.20 \\ 
SDSS J204524+150825  & 4 &61  &5136&32   &2.7&0.38 &-3.8&0.02 & 0.3  &0.28 \\ 
SDSS J231027+231003  & 7 &62  &6213&20   &4.4&0.05 &-3.8&0.04 & 1.5  &0.08\\  
SDSS J222505+164322  & 6 &50  &5969&21   &4.3&0.06 &-3.4&0.04 & 0.8  &0.39 \\ 
\hline
\hline              
HE 0233$-$0343$^{(1)}$       &6&65 &6150 &16 &4.9&0.02& -4.0 &0.03&  2.2 &  0.15  \\
SDSS J1029+1729$^{(2)}$      &6&69&5845&13&5.0&0.01&  -4.4&0.04&   $0.4^b$&   0.75  \\
SDSS J1313$-$0019$^{(3,4)}$  &7&66&5525&17&3.6&0.05& -4.7&0.03&   2.8&   0.03  \\
HE 1327$-$2326$^{(5,6)}$     &2&133&6101&11&4.3&0.02&  -4.8&0.03&   3.0&   0.03   \\
G64$-$12$^{(6)}$             &2&321&6377&5&4.8&0.01&  -3.2& 0.01&  $0.4^b$&  0.15   \\
SDSS  J1442$-$0015$^{(7)}$   &7&42&6036&31&4.9&0.08&  -4.4&0.01&   $0.5^b$&  0.47   \\
\hline              
\end{tabular}
\end{center}
 \tablefoot{ $^a$ Signal-to-noise ratios have been calculated as 
 the average of S/N values for the  entire SDSS spectrum.
 $^b$ Values below our limit detection.}

\tablefoot{$\Delta$ is the internal uncertainty of the parameters derived with \ferre\,.}
\tablefoot{$(1)=$\cite{han14}; $(2)=$\cite{caff12I}; $(3)=$\cite{alle15}; $(4)=$\cite{fre15};
$(5)=$\cite{fre05}; $(6)=$\cite{aoki06I}; $(7)=$\cite{caff13II}}
\end{table*}
\subsubsection{SDSS J1442$-$0015}\label{w2909}

The S/N ratio of this spectrum is the lowest in our sample since it is one 
of the faintest metal-poor star in the $\left[{\rm Fe/H}\right]<-4.0$ regime. 
Even so, the original set parameters derived from UVES spectroscopy
by \citet{caff13II}, $T_{\rm eff}=5850\pm 150\,K$, $\logg=4.0\pm 0.5$, and 
$\left[{\rm Fe/H}\right]=-4.1\pm 0.2$, are in fair agreement with 
our determinations from ISIS data:  
$T_{\rm eff}=6036\pm 102\,K$, $\logg=4.9\pm 0.5$, and 
$\left[{\rm Fe/H}\right]=-4.4\pm 0.2$.

\begin{figure*}
\begin{center}
{\includegraphics[width=170 mm]{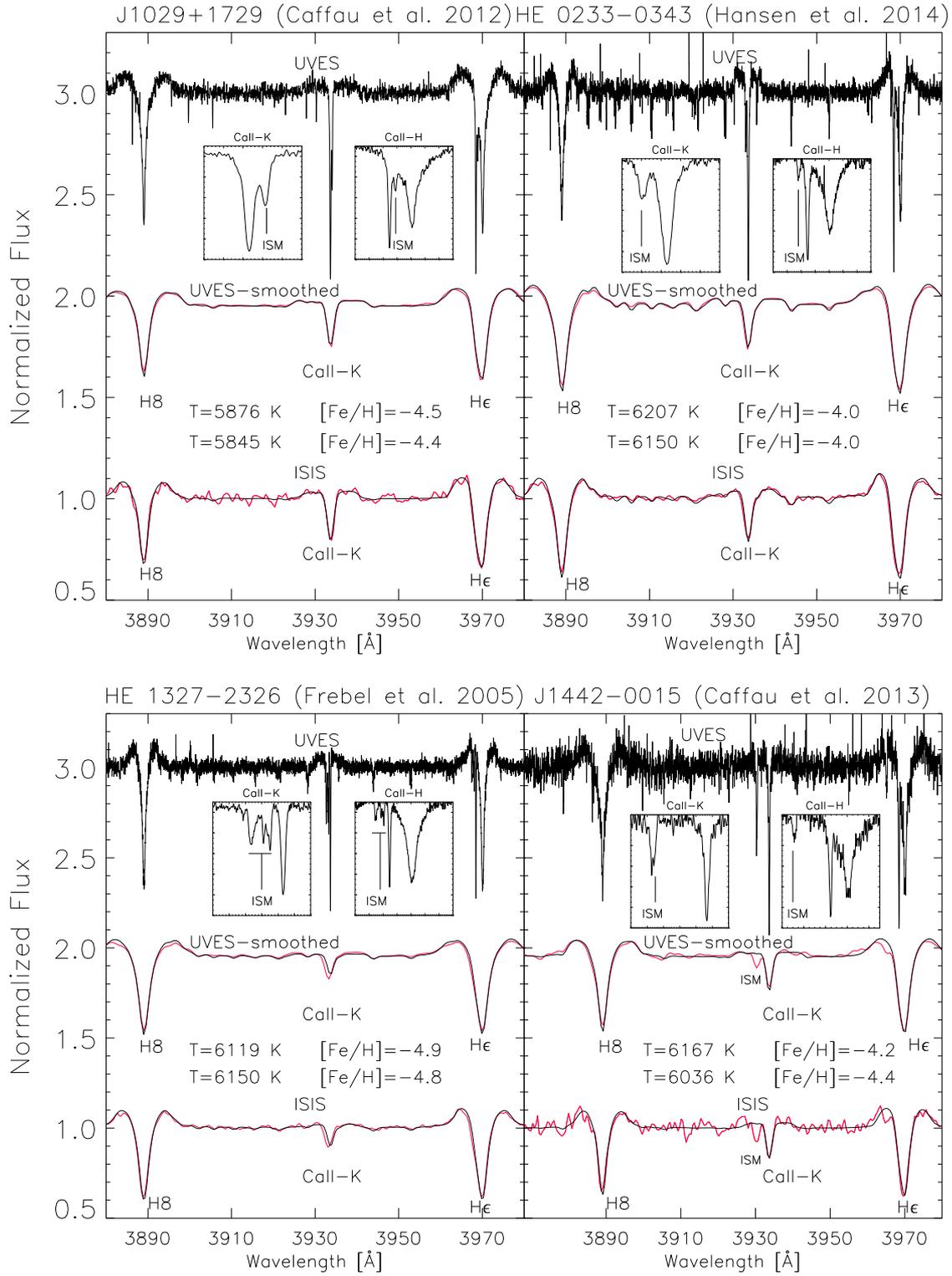}}
\end{center}
\caption{Spectra in the range of the \ion{Ca} {II} lines for four well-known 
metal-poor stars. 
From top to bottom: UVES, UVES smoothed to ISIS resolution, ISIS spectra (black line), and the best medium-resolution fit (red line).
Details of the calcium lines in the UVES spectra are included in the insets 
together with the effective temperature and metallicity derived with FERRE for 
UVES-smoothed and ISIS spectra.}
\label{comp}
\end{figure*}

The spectrum of J1442$-$0015  clearly shows an ISM contribution to 
the observed calcium absorption, but this component is already resolved 
in the ISIS data (see Fig.~\ref{comp}). 
The effective temperature  adopted by \citet{caff13II}, which is slightly lower 
than our \ferre\ estimate, should lead to a difference of at least $\sim$0.4\,dex  
between the two metallicity determinations but in the opposite direction to what 
we find. In addition, our analysis of the UVES spectrum smoothed to the
resolution of ISIS provides a $T_{\rm eff}=6167\,K$,
$\logg=4.9$ and $\left[{\rm Fe/H}\right]=-4.2$, consistent with the results
from ISIS.

\begin{figure*}
\begin{center}
{\includegraphics[width=180 mm]{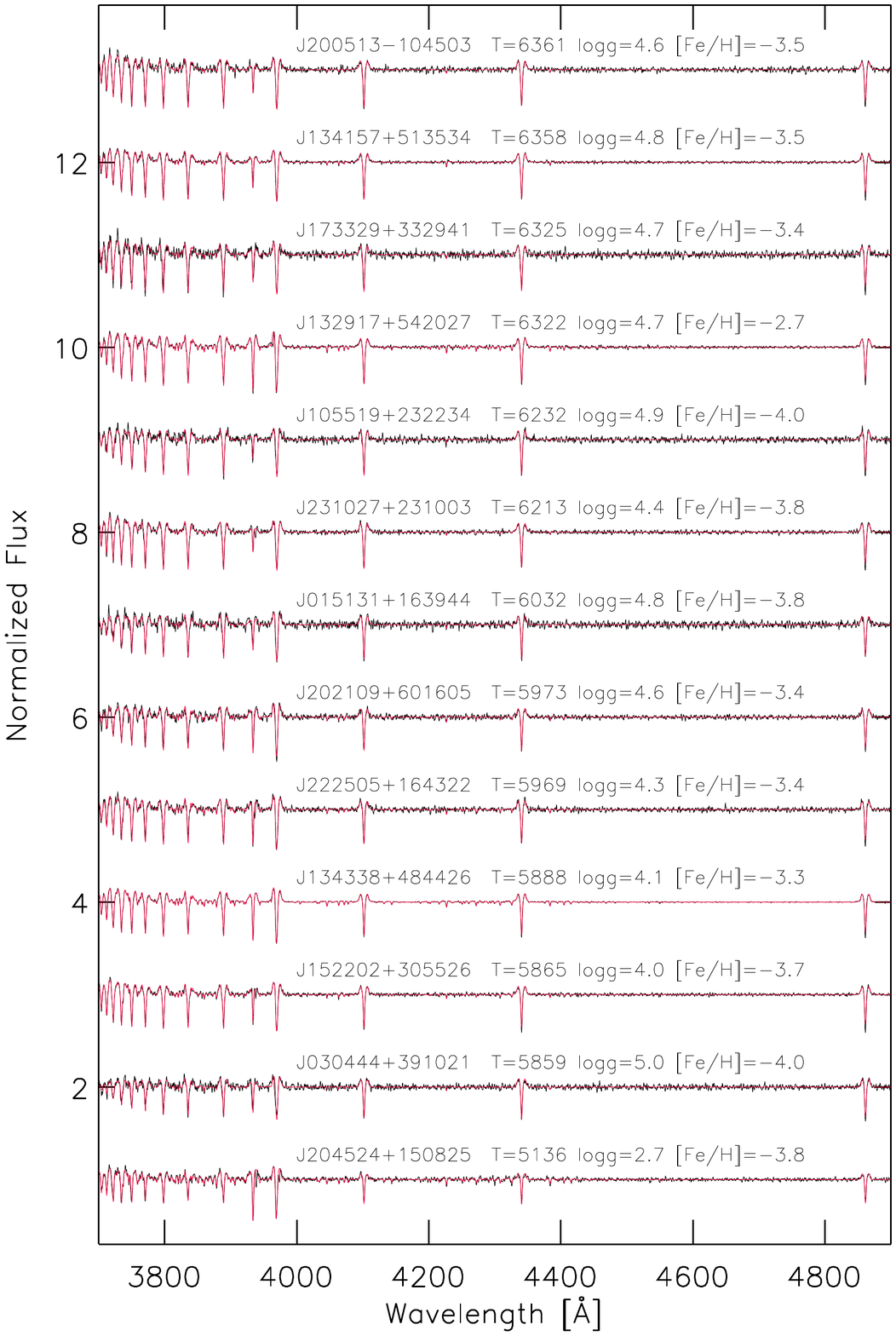}}
\end{center}
\caption{ISIS/WHT blue arm spectra (3700\,\AA-4900\,\AA) for the full sample
(black line) and the best fit calculated with FERRE (red line). The main 
stellar parameters are also displayed.}
\label{all}
\end{figure*}

%\subsubsection{Systematic differences between analyzes}
%\label{sys}

%\textbf{Comparing our results with the three reference objects 
%observed in high resolution by \citet{aoki06I},  \citet{fre15},
%and \citet{fre05} (see sections \ref{g64-12},\ref{allende},\ref{frebel}) 
%we obtained the following differences:
%139\,K ($\sigma=164$ K) in $T_{\rm eff}$, 0.8\,dex $\sigma=0.3$ in $\log g$,
%0.15 dex ($\sigma=0.21$) in [Fe/H]\footnote{The HE 1327$-$2326 metallicity 
%does not contribute to this comparation cause the ISM as explained in \ref{frebel}}, and
%0.24 dex ($\sigma=0.22$) in [C/Fe].
%For the two stars observed by \citet{caff13II} 
%and \citet{caff11}  (see sections \ref{caffau},\ref{w2909}), we find differences of 
%110 K ($\sigma=107$) K in $T_{\rm eff}$,  0.95 dex ($\sigma=0.07$) in $\log g$, 
%and 0.3\,dex in [Fe/H]\footnote{The SDSS J1029+1729 metallicity 
%does not contribute to this comparation cause the ISM as explained in \ref{caffau}}
%}.

\subsection{New set of extremely metal-poor stars}\label{new}

Following the same methodology as in Section~\ref{known}, we analyzed spectra
from the ISIS instrument for a sample of metal-poor star candidates identified
from SDSS and LAMOST.
The mean S/N of these spectra is around 75, so we expect the \ferre\ results 
to be reliable.
Our effective temperatures are trustworthy, since we were able to 
recover the temperature for several well-known metal-poor stars whose metallicities 
are consistent with the literature values. In addition, the derived metallicities 
by \ferre\ are also
reliable and consistent with the results obtained from the
analysis of SDSS spectra~\citep[see][for further details]{alle14}.
Only our surface gravities appear to be subject to a systematic error of 
about $0.5$ dex. 
This is in contrast to the small random uncertainties we derive for 
our stellar parameters, and to the surface gravity in particular (Sect. \ref{anal}):
 lower than 0.1 dex at S/N$\sim 50$.

\begin{figure}
\begin{center}
{\includegraphics[width=90 mm, angle=180]{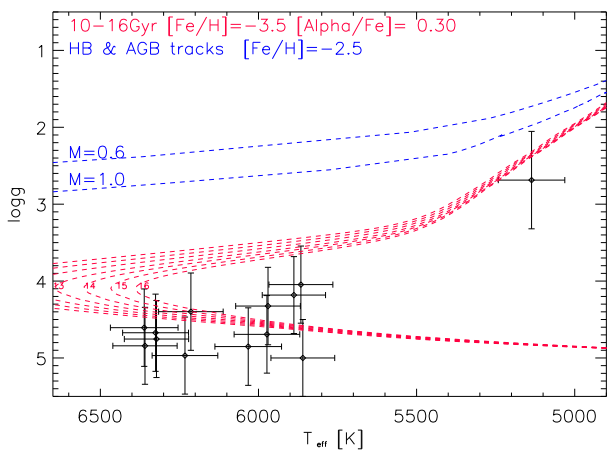}}
\end{center}
\caption{DARMOUTH isochrones for $\left[{\rm Fe/H}\right]=-3.5$ and different 
ages from 16 to 10 Gyr (red dashed lines), and blue dashed lines are the HB and AGB 
theoretical tracks for $\left[{\rm Fe/H}\right]=-2.5$ for two different 
relative masses ($M=0.6$ and $M=1.0$). The black diamonds represent the 
stars of this work and their internal uncertainties derived with
the FERRE code.}
\label{iso}
\end{figure}

Figure~\ref{iso} shows DARMOUTH isochrones \footnote{The Dartmouth Stellar 
Evolution Program (DSEP) is available from www.stellar.dartmouth.edu}, 
HB,  and AGB tracks compared to the stellar parameters derived using \ferre\ 
on the ISIS spectra and their derived error bars.
Our mean uncertainty in the metallicity determination is 0.12\,dex, 
whereas the mean uncertainty in $T_{\rm eff}$ is 103\,K. 

\begin{figure}
\begin{center}
{\includegraphics[width=100 mm, angle =180]{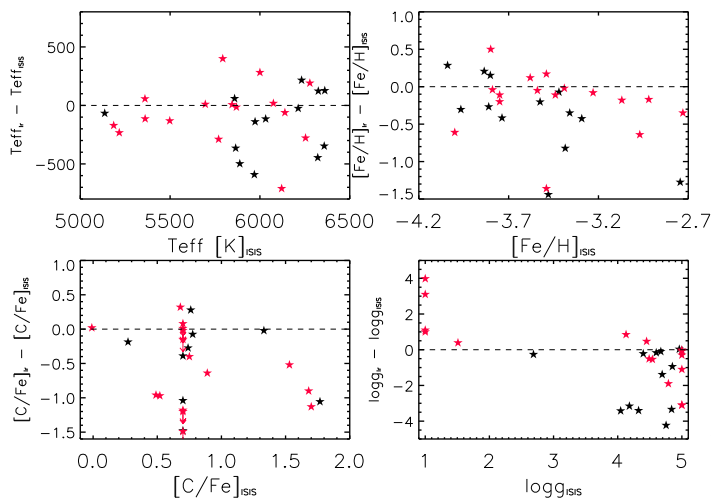}}
\end{center}
\caption{Comparison between the main stellar parameters derived with FERRE from low-resolution
(SDSS and LAMOST) data and from ISIS spectra.
Objects included in our sample appear as black filled stars while objects from 
our previous work \citet{agu16} are red filled stars.
}
\label{newplot}
\end{figure}

The mean metallicity difference between our first metallicity estimates
from SDSS/LAMOST spectra and the second estimate we obtain from higher quality 
ISIS spectra is 0.31\,dex with a standard deviation of 0.20\,dex. 
In this computation we excluded the stars J134157+513534 
and J132917+542027, since the metallicity differences for them are much larger.
The LAMOST spectra of these stars show several artifacts that do 
not allow us to perform a correct continuum normalization with our 
running-mean algorithm, and consequently, we cannot derive reliable parameters from these data.
However, our ISIS spectra have significantly higher quality, 
S/N$\sim 82$ and $76$, and we finally derived a  reliable metallicity 
for both stars, [Fe/H]$=-2.7$ and [Fe/H]$=-3.5$, respectively.
In Fig. \ref{newplot} we compare the results that are summarized in Table \ref{AnalysisResults} with those from the
low-resolution analysis (Table \ref{basic}). In addition, we reanalyzed the stars studied in \citet{agu16}
with the improved methodology we presented here.
We obtain a slightly higher dispersion for candidates with lower quality
spectra.

Our analysis uncovers two objects (J0304+3910 and J1055+2322) 
at [Fe/H]$\simeq-4.0$  and six more at $-3.5\geq$[Fe/H]$>-4.0$, in 
the domain of the extremely metal-poor stars, and most of them 
appear to be dwarfs at $\logg \ge 4.0$.
A deeper study of this sample using high-resolution spectroscopy 
would provide abundances for additional elements, which would help to constrain 
the nature of the stars and the early chemical evolution of the Galaxy.

\subsection{Carbon abundances}\label{carbon}

Carbon-enhanced metal-poor (CEMP) stars are defined by \citet{bee05} 
as stars with [C/Fe]$\geq +1.0$. The fraction of CEMP/EMP stars 
increases as metallicity decreases. Deriving reliable carbon abundances 
in metal-poor stars using medium-resolution spectroscopy is not always 
possible \citep{boni15}.
For the well-known EMP sample studied in Section~\ref{known}, we 
recover carbon abundances that are compatible with the literature values for 
three CEMP stars: SDSS J1313$-$0019, HE 1327$-$2326, 
and HE 0233$-$0343  (see Table 3).

\begin{table}
\caption{Carbon abundance results.}      % title of Table
\label{car_tab}      % is used to refer this table in the text
\centering                          % used for centering table
\begin{tabular}{c c c c}        % centered columns (4 columns)
\hline\hline                 % inserts double horizontal lines
star & Bibliography & This work\\
&UVES/MIKE&ISIS\\    % table heading 
&A(C)($\Delta)$&A(C)($\Delta)$\\    % table heading 
\hline                        % inserts single horizontal line
  HE 0233$-$0343  & 7.18(0.24) & 6.7(0.3) \\      % inserting body of the table
   SDSS J1029+1729 & n.d. &4.5(0.8)      \\
    SDSS J1313$-$0019& 6.46(0.28) & 6.6(0.3)     \\
   HE 1327$-$2326 & 7.20(0.2) & 6.7(0.3)     \\
    G64$-$12&  5.8(0.2) &5.7(0.4)   \\ 
   SDSS  J1442$-$0015 & n.d. &4.6(0.6)      \\ 
\hline   %inserts single line
%\tablefoot{The used $\rm A(C)_{\odot}$=8.50.}

\end{tabular}
\end{table}

\begin{figure}
\begin{center}
{\includegraphics[width=90 mm, angle=180]{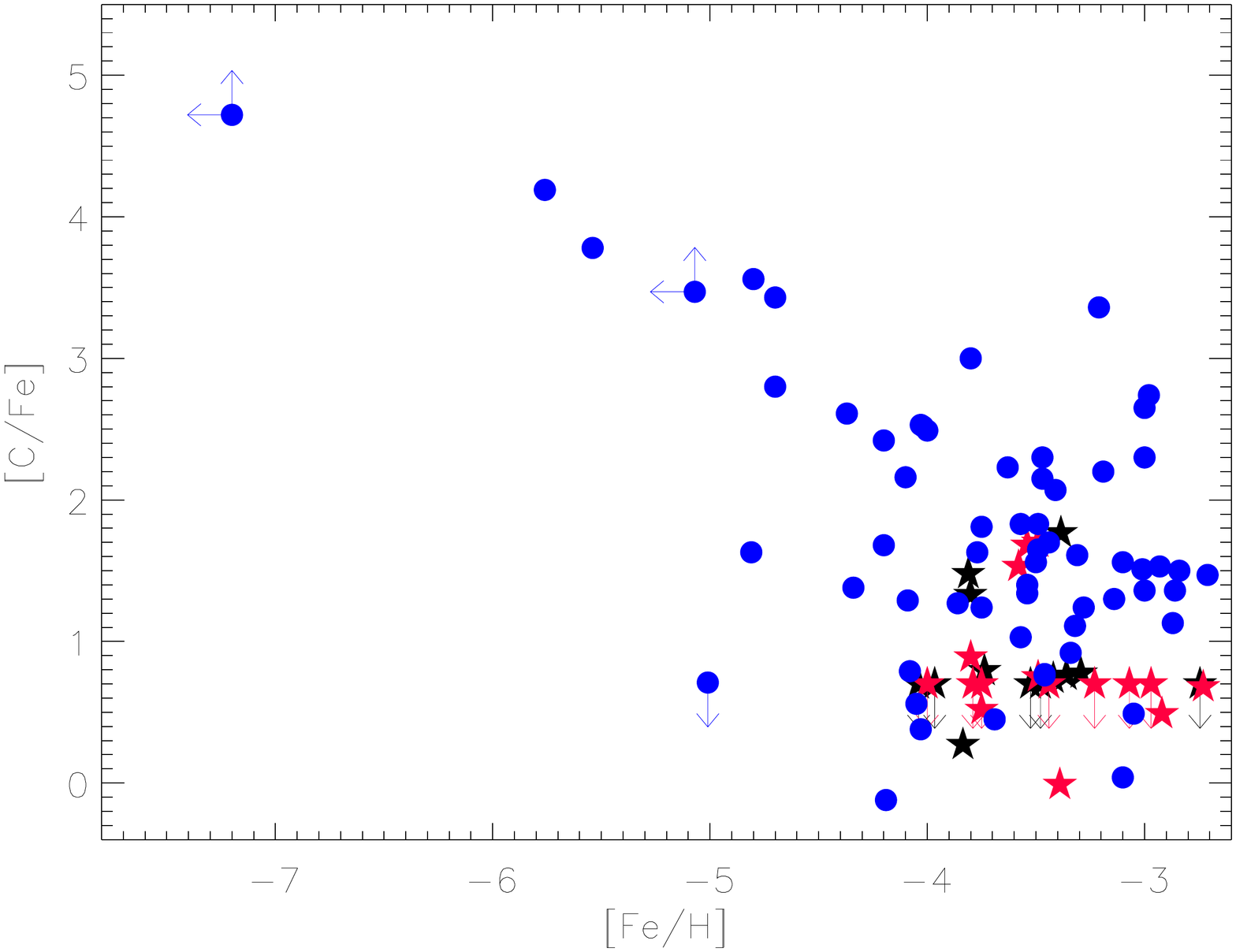}}
\end{center}
\caption{Carbon over iron versus metallicities of stars presented in this work (black filled stars) 
and those from \citet{agu16} (red filled stars) analyzed with 
the improved methodology, 
together with stars from the literature  
 \citep{siv06,yong13II,fre05,fre06,caff14,alle15} 
represented by blue filled circles.}
\label{carbon2}
\end{figure}

Figure \ref{carbon2} shows the carbon abundances derived from ISIS spectra for the targets studied in this work,
together with those from \citet{agu16}, with the improved methodology, and other
 CEMP stars from the literature.
The data suggest that the lower the metallicity, the stronger the carbon enhancement, as expected (see, e.g., \citet{lee13}).
Our ability to measure carbon abundances decreases as the effective 
temperature of the star increases because of CH dissociation. 
Empirically, to measure carbon abundances at the level of [C/Fe]$\geq +0.5$, the star should have
 $T_{\rm eff}\sim5500/5700$\,K. Moreover, if the effective temperature lies at $\sim5000/5100$\,K, we 
are able to measure carbon at  [C/Fe]$\geq +0.0$.
In addition, an S/N of at least 30 is required to derive reliable carbon abundances.  
Table \ref{AnalysisResults} summarizes our results and shows that 
our sample includes at least three confirmed 
CEMP stars: SDSS J0151+1639, SDSS J1733+3329, and 
SDSS J2310+1210 with [Fe/H]$=-3.8$, $-3.4$, and $-3.8$, respectively, 
in perfect agreement with the statistics by \citet{cohen09,boni15}.
In addition, five more stars appear to be very close to 
our detection limit of [C/Fe]$\sim+0.7-0.8$.

\section{Conclusions}\label{discuss}

We have demonstrated that our methodology for identifying metal-poor 
stars using low- and medium-resolution spectroscopy is effective. 
We were able to scrutinize  more than 
2.5 million spectra from SDSS and LAMOST 
to identify extremely metal-poor star candidates, which we 
followed up with significantly higher S/N and slightly 
higher resolution ISIS observations.
Our success in identifying metal-poor candidates is quite high, 
with only one star, J1329+5420, at [Fe/H]$>-3$.

The use of the \ferre\ code and a custom  grid of synthetic spectra
allowed us to simultaneously derive  the atmospheric parameters 
and the carbon abundance, which improved our values from SDSS or LAMOST.
We showed that \ferre\ can be used efficiently 
on extremely metal-poor stars.
We make our tools publicly available to
facilitate the cross-calibration of results from other teams. 
Our method offers an excellent way to 
identify and analyze CEMP star candidates without the need for 
high-resolution spectroscopy. We presented a new EMP sample and 
reliable determinations of the metallicities and carbon abundances of these stars.

These stars, especially the two at [Fe/H]$<-4.0$, 
are good candidates for follow-up high-resolution observations.
This domain is sparsely populated, with fewer than 30 known stars, and it is of
high interest for investigating the early chemical evolution of the Galaxy.
The fact that we have used a grid of synthetic spectra including carbon 
as a free parameter not only in the synthesis but in the model atmospheres 
helps us to detect promising extremely metal-poor stellar 
candidates and derive their carbon abundances when 
[C/Fe]$>0.7-1.0$ and S/N$>30/40$, and even lower carbon enhancements and/or
$S/N$ values if the star is colder than 5500\,K.

We included in our work several 
well-known metal-poor stars that were previously analyzed in the literature. 
A  study of these objects is useful to check for systematic 
differences that are due to multiple analysis methods.
Our results shown that we are able to recover the effective temperature 
with an uncertainty of about 100\,K and the metallicity within 0.2\,dex. 
In addition, the comparison of our parameters with model
isochrones suggests that our surface gravities have a systematic uncertainty
of about 0.5 dex. Our study also confirms that we are able to recover the 
carbon abundance for CEMP stars using medium-resolution spectra.

Future work should include observations at higher resolution of the most 
interesting extremely metal-poor stars identified in this paper
to study their chemical abundance patterns in detail.

\begin{acknowledgements}
DA is thankful to the Spanish Ministry of Economy and Competitiveness 
(MINECO) for financial support received in the form of a 
Severo-Ochoa PhD fellowship, within the Severo-Ochoa International PhD 
Program.
DA, CAP, JIGH, and RR acknowledge the Spanish ministry project 
MINECO AYA2014-56359-P. 
JIGH acknowledges financial support from the Spanish Ministry of 
Economy and Competitiveness (MINECO) under the 2013 Ram\'on y Cajal 
program MINECO RYC-2013-14875. 

This paper is based on observations made with the William Herschel
Telescope, operated by the Isaac Newton Group at the Observatorio del 
Roque de los Muchachos, La Palma, Spain, of the Instituto de 
Astrof{\'i}sica de Canarias. We thank the ING staff members for their 
assistance and efficiency during the four observing runs in visitor mode.\\
The author gratefully acknowledges the technical expertise 
and assistance provided by the Spanish Supercomputing Network 
(Red Espanola de Supercomputacion), as well as the computer resources used:
the LaPalma Supercomputer, located at the Instituto de Astrofisica de Canarias.\\
\end{acknowledgements}


\bibliography{biblio}

\end{document}